\documentclass[runningheads]{comsis2}

\usepackage{amsmath}
\usepackage{amssymb}
\usepackage{wasysym}
\usepackage{textcomp}
\usepackage{graphicx}
\usepackage{siunitx}
\DeclareSIUnit \rpm {\text{rpm}}
\DeclareSIUnit \g {\text{$g$}}
\DeclareSIUnit \Molar {\textsc{m}}
\usepackage[version=4]{mhchem}

\usepackage{array}
\usepackage{dcolumn}  
\usepackage{rotating}
\usepackage{ragged2e}
\usepackage[greek,english]{babel}
\usepackage{lmodern}      
\usepackage{setspace}
\usepackage{textgreek}
\usepackage[labelfont=bf]{caption}
\usepackage{lscape}
\usepackage{multicol}
\usepackage{multirow}
\usepackage{booktabs}
\usepackage{tabularx}
\usepackage{color}
\usepackage{xcolor}
\usepackage{colortbl}
\usepackage{aeguill}
\usepackage{soul}
\usepackage[activate=true,tracking=true,factor=900,stretch=7,shrink=7]{microtype}
\usepackage{hyperref}
\hypersetup{
  colorlinks = false,
  linkbordercolor = {white},
  allbordercolors = {lightgray}
}

\usepackage{orcidlink}
\usepackage{url}
\usepackage{xurl}   

\newcolumntype{L}[1]{>{\raggedright\arraybackslash}p{#1}}  
\newcolumntype{C}[1]{>{\centering\arraybackslash}p{#1}}    
\newcolumntype{R}[1]{>{\raggedleft\arraybackslash}p{#1}}   

\newcommand{\eg}{e.\medspace g.}
\newcommand{\ie}{i.\medspace e.}

\newcommand{\gl}{\guillemotleft}
\newcommand{\gr}{\guillemotright}

\def\journalissue{Computer Science and Information Systems 00(0):0000--0000}
\def\paperidnum{https://doi.org/10.2298/CSIS123456789X}
\title{RAMSES: Secure high-performance computing for sensitive data}

\titlerunning{RAMSES: Next-generation HPC security}

\author{Peter Heger\inst{1}\orcidlink{0000-0003-2583-2981} \and Lech Nieroda\inst{1}\orcidlink{0000-0001-9745-8883} \and Roland Pabel\inst{1}\orcidlink{0009-0008-0013-067X} \and Christoph Stollwerk\inst{1}\orcidlink{0009-0007-2529-0314} \and Stefan Borowski\inst{1}\orcidlink{0009-0006-8126-6434} \and Kamil Tokmakov\inst{2}\orcidlink{0000-0002-3160-2309} \and Michael Commer\inst{1}\orcidlink{0000-0003-0015-9217} \and Martin Peifer\inst{3}\orcidlink{0000-0002-5243-5503} \and Stefan Wesner\inst{1}\orcidlink{0000-0002-7270-7959} \and Viktor Achter\inst{1}\orcidlink{0000-0002-3813-0746}}

\authorrunning{Heger et al.}

\institute{IT Center Cologne (ITCC)\\
   University of Cologne\\
   Weyertal 121\\
   50931 Cologne, Germany\\
   \email{heger@uni-koeln.de},
   \email{nieroda@uni-koeln.de},
   \email{pabel@uni-koeln.de},
   \email{stollwerk@uni-koeln.de},
   \email{borowski@uni-koeln.de}, 
   \email{commer@uni-koeln.de}, 
   \email{wesner@uni-koeln.de},
   \email{vachter@uni-koeln.de}
   \and
   NEC Deutschland GmbH\\
   Fritz-Vomfelde-Straße 14-16\\
   40547 D\"usseldorf, Germany\\
   \email{kamil.tokmakov@emea.nec.com} 
   \and 
   Department of Translational Genomics\\
   University of Cologne\\
   50931 Cologne, Germany\\
   \email{mpeifer@uni-koeln.de}
}

\begin{document}

\maketitle

\begin{abstract}
Traditionally, the architecture of high-performance computing (HPC) systems is tailored for speed, while highly secure computer systems must sacrifice speed for security. 
However, a wide range of scientific domains, such as the life sciences, call for a combination of performance and security to allow processing sensitive data at scale. 
Here, we present RAMSES (Research Accelerator for Modeling and Simulation with Enhanced Security), an HPC system designed from the ground up to deliver high performance within a robust security framework. 
RAMSES integrates hardware-based memory encryption of AMD processors with state-of-the-art file encryption from IBM Storage Scale and the Thales CipherTrust manager, establishing an HPC platform that ensures continuous encryption throughout the data life cycle---at rest, in transit, and in use---in compliance with major data protection standards (European General Data Protection Regulation, ISO\slash\allowbreak IEC 27001 certification, and Federal Information Processing Standards). In addition, we implemented advanced operating system hardening, a multi-layered security architecture, and mandatory multi-factor authentication to adapt the HPC environment to increased security demands. 
Benchmark results from the biomedical sector demonstrate that the performance impact of the secure environment is limited and that integration of the conflicting requirements speed and security can be achieved while preserving a coherent, flexible, and user-friendly system.

\vspace{6pt}\textbf{Keywords:} Secure high-performance computing, confidential computing, end-to-end encryption, memory encryption, Hardware Security Module (HSM), AMD SEV/SME, Trusted Research Environment (TRE), Thales, multi-factor authentication (MFA), Information Security Management System (ISMS), key management, operating system hardening, threat model, secure boot, SR-IOV, SPANK. 
\end{abstract}

\section{Introduction}
HPC systems are typically composed of numerous individual compute nodes and storage devices, coupled through high-speed interconnects, and are widely seen as drivers of scientific discovery and innovation \cite{NSF2015,Schlick2021,Xia2022,Chen2024}. 
The performance of the fastest supercomputers has reached the exascale level, with the execution of more than one exaFLOP/s (floating point operations per second) on the latest German supercomputer, Jupiter (\url{https://www.fz-juelich.de/de/jupiter}), the 4th fastest computer in the world (as of November 2025; \url{https://top500.org/lists/top500/2025/11/}). 
While access to supercomputers is restricted and users are generally not allowed to perform system-critical tasks, such as installing and modifying system software, conventional HPC systems rely on Linux's standard security mechanisms which often lack advanced features for system and\slash\allowbreak or data security \cite{CSA2020,Nist800-223}. 

In contrast, highly secure computer systems, \eg\/ from the banking or military sector, are typically small in comparison to TOP500-listed systems (\url{https://top500.org/}). Their design follows official security recommendations and often involves dedicated hardware and software layers on top of OS (operating system) level protection, for example HSMs (hardware security modules), TPMs (trusted platform modules), secure boot, tamper detection (hardware level), and specific software solutions that address different aspects of cybersecurity, such as key management systems, network security tools, or monitoring solutions \cite{Nist800-207,Gaddam2021,CRI2023}. Many of these measures are considered mandatory to ensure data integrity, transaction throughput, or regulatory auditability in their specific contexts \cite{FSB2017,NIST800-53r5,DoD2022,Hussain2025_airgap}. 

According to NIST (National Institute of Standards and Technology) principles, HPC system architecture can be divided into several zones, such as access, compute, management, and storage. Each zone is exposed to distinct threat classes. They include denial-of-service attacks, perimeter scanning, and traffic interception in the access zone; side-channel and co-tenancy attacks in the compute zone; privilege escalation in the management plane; and data confidentiality or integrity violations in the storage layer \cite{Nist800-223}. 
Consequently, the design of secure HPC architectures requires an in-depth approach that reduces zone-specific threats through appropriate countermeasures, including encryption of data at rest, in transit, and in use; strong identity and access management; comprehensive audit and logging mechanisms; compute and workload isolation; network segmentation; and hardware- and software-based cryptographic protection. 

RAMSES addresses the above-mentioned security threats. Specifically designed for secure data processing, RAMSES integrates hardware-enforced memory encryption, cryptographically protected storage, and systematic operating-system hardening, while preserving usability for scientific workloads.

\section{State of the Art}
\subsection{Secure Environments for Scientific Computing} 
Implementations of secure computing environments cover a broad spectrum, from regulatory compliance frameworks to hardware-enforced protection of data in use. At the basic level, many HPC centres adopt formal certification standards such as ISO\slash\allowbreak IEC~27001, which define requirements for an information security management system (ISMS), including risk assessment, access control, incident response, and continuous auditing \cite{ISO27001}. The use of validated cryptographic modules is often an additional requirement. For example, the U.S.\@ FIPS~140-2 and FIPS~140-3 standards specify security requirements and approved algorithms for cryptographic modules, which are independently evaluated under the Cryptographic Module Validation Program (CMVP) and assigned assurance levels based on physical and logical protections \cite{NIST-FIPS140-2_upd2,NIST-FIPS140-3,ISO_IEC_19790_2025}. These certifications warrant process maturity and cryptographic correctness, but do not require specific technical safeguards such as encryption of data, tenant isolation, or protection against privileged insider threats. 
While a growing number of academic HPC centres in Europe and North America operate under such certification programs (Table~\ref{certified_HPC_systems}), certified HPC systems often still adhere to conventional architectures and remain unsuitable for sensitive workloads. 

Trusted Research Environments (TREs) represent a more mature level of security. TREs are controlled computing environments explicitly designed for sensitive or regulated data and typically align with established frameworks such as the Five Safes model for safe research access to data \cite{desai2016fivesafes}. They combine strong identity and access management, restricted network connectivity, monitoring and audit logging, and user training to reduce risks across access, compute, management, and storage zones \cite{tre_green_paper2020,Kavianpour2022}. Architectural guidance and operational practices for TREs are reliable and widely adopted, particularly in the biomedical and social sciences. As of late 2024, more than 40 TREs have been identified across Europe \cite{EGI2024_tre_landscape}, including HPC-enabled platforms at Cambridge, the University of Edinburgh, or Imperial College London (Table~\ref{certified_HPC_systems}). However, TREs primarily rely on software isolation and organisational controls and typically do not protect data in use or at rest. 

Further improvements in protection are achieved through Trusted Execution Environments (TEEs), which introduce hardware-enforced isolation mechanisms that protect code and data even from privileged system software \cite{globalplatform_tee_spec}. TEEs address threats explicitly highlighted in NIST SP~800-223, such as memory disclosure, remapping, and privileged access in the compute zone, by providing encryption and integrity protection for data in use, \ie\/, memory encryption \cite{Nist800-223}. While TEEs are increasingly deployed in public cloud infrastructures, their integration into shared, large-scale HPC environments remains limited due to operational complexity and performance considerations. Nevertheless, recent attempts to operate TEEs in HPC context report acceptable performance losses and thus encourage the productive use of TEEs, for example for bioinformatics workloads \cite{Akram2021,Kessler2025,Dokmai2025,Coppolino2025}. 

An intermediate approach bridging conventional HPC and hardware-backed confidential computing was proposed by Nolte et al.\@ \cite{Nolte2022}. Their secure HPC workflow combines encrypted containers and datasets, centralised key management via HashiCorp Vault (\url{https://www.vaultproject.io/}), signed job scripts, and strict node and network isolation. Data is encrypted prior to upload and decrypted only within isolated execution environments on the cluster, supporting multi-node workflows and parallel encrypted I/O. Although this approach substantially improves the protection of data at rest and in transit, it introduces significant complexity for users and does not address threats to data in use, such as physical memory extraction or privileged RAM inspection. 

In summary, existing secure HPC solutions demonstrate increasing levels of protection, from organisational certification through TRE-style controlled environments to hardware-enforced confidential execution in TEEs, but no single approach yet provides comprehensive, zone-spanning security as recommended by NIST SP~800-223. Therefore, achieving integrated protection for data at rest, in transit, and in use within shared HPC systems remains an open challenge.

\setlength{\tabcolsep}{7pt}  

\begin{table}[hbt!]
\centering
\footnotesize
\begin{tabular}{@{}p{5.1cm} p{2.1cm} p{1.5cm} l l@{}}
\toprule
\textbf{Site} & \textbf{Certification} & \textbf{Extras} & \textbf{Link} & \textbf{Year} \\
\midrule

HLRS (H\"ochstleistungs\-rechenzentrum Stutt\-gart), University of Stuttgart
  & ISO27001
  & ---
  & \href{https://www.hlrs.de/news/detail/hlrs-achieves-iso-27001-certification-for-information-security-management}{a}
  & 2023
  \\\\\addlinespace[-0.6em]

Secure Research Computing Platform (SRCP), University of Cambridge
  & ISO27001, NHS Data Security and Protection Toolkit
  & encryption, monitoring
  & \href{https://www.hpc.cam.ac.uk/secure-research-computing}{b}
  & 2016
  \\\\\addlinespace[-0.6em]

SDU eScience Center, University of Southern Denmark
  & ISO27001
  & ---
  & \href{https://escience.sdu.dk/index.php/organization/iso-27001/}{c}
  & 2020
  \\\\\addlinespace[-0.6em]

Secure Research Infrastructure (Citadel), Princeton University
  & NIST SP 800-171, NIST SP 800-53, HIPAA
  & encryption, MFA, monitoring
  & \href{https://researchcomputing.princeton.edu/systems/secure-research-infrastructure}{d}
  & 2021
  \\\\\addlinespace[-0.6em]

Centre for Advanced Research Computing, University College London
  & ISO27001, NHS Data Security and Protection Toolkit
  & TRE, MFA, Data Security Awareness course
  & \href{https://www.ucl.ac.uk/advanced-research-computing/sensitive-data-and-trusted-research-environments}{e}
  & 2014
  \\\\\addlinespace[-0.6em]

GWDG, G\"ottingen University
  & ISO9001, ISO27001
  & Secure HPC workflow \cite{Nolte2022}
  & \href{https://docs.hpc.gwdg.de/services/secure-hpc/index.html}{f}
  & 2023
  \\\\\addlinespace[-0.6em]

SURF (Samenwerkende Universitaire Rekenfaciliteiten) Research Cloud
  & ISO27001
  & ---
  & \href{https://www.surf.nl/en/services/compute/surf-research-cloud}{g}
  & 2020
  \\\\\addlinespace[-0.6em]

Edinburgh Parallel Computing Centre (EPCC), University of Edinburgh
  & ISO27001, NHS Data Security and Protection Toolkit
  & TRE
  & \href{https://www.epcc.ed.ac.uk/hpc-services/trusted-research-environments}{h}
  & 2018
  \\\\\addlinespace[-0.6em]

Big Data and Analytical Unit (BDAU), Imperial College London
  & ISO27001, NHS Data Security and Protection Toolkit
  & TRE
  & \href{https://www.imperial.ac.uk/admin-services/ict/self-service/research-support/rcs/service-offering/big-data-and-analytical-unit-bdau/}{i}
  & 2017
  \\\\\addlinespace[-0.6em]
  
Computational Research, Engineering and Technology Environment (CREATE), King's College London
  & ISO27001, NHS Data Security and Protection Toolkit
  & TRE, MFA, encryption
  & \href{https://docs.er.kcl.ac.uk/CREATE/TRE/tre/}{j}
  & 2021
  \\\\\addlinespace[-0.6em]

CINECA (Consorzio Interuniversitario per il Calcolo Automatico dell'Italia Nord Orientale)
  & ISO9001, ISO27001, ISDP 10003
  & ---
  & \href{https://www.cineca.it/en/about-us/cineca-today/certifications}{k}
  & 2022
  \\\\\addlinespace[-0.6em]

Advanced Research Computing, Cardiff University
  & ISO27001, ISO9001
  & ---
  & \href{https://www.cardiff.ac.uk/news/view/2966225-research-computing-moves-to-new-data-centre}{l}
  & 2025
  \\
  
\addlinespace[0.1em]
\bottomrule
\end{tabular}

\caption{\textsf{\textbf{List of certified HPC centers in North America and Europe.} 
Measures that improve security beyond formal certification are listed in column \gl Extras\gr\/. MFA: Multi-factor authentication; TRE: Trusted Research Environment; TEE: Trusted Execution Environment. For web links to the respective centers, use the hyperlinks provided in column \gl Link\gr\/. Please note that this list contains a selection of highly visible, certified HPC centers, but may be incomplete. }}
\label{certified_HPC_systems}
\end{table}

\subsection{Encryption Techniques: Symmetric, Asymmetric, Hybrid, Homomorphic}
Encryption is a central component of any secure computing environment. Symmetric encryption uses a shared secret key for both encryption and decryption. Its classic implementation is the Advanced Encryption Standard (AES), specified in FIPS publication 197 \cite{FIPS197}. 
Symmetric encryption methods work with fixed-size data blocks using simple deterministic transformations, which can be efficiently parallelized and are widely supported by dedicated CPU instructions (\eg\/, AES-NI). 
As a result, symmetric encryption achieves high throughput and low latency, making it well-suited for bulk data protection, such as encrypting entire file systems or object-store volumes. Owing to its performance characteristics, symmetric encryption is the standard mechanism for protecting data at rest and data in transit, \eg\/, within the protocol for transport layer security (TLS). 

Asymmetric encryption (public-key cryptography) relies on a key pair consisting of a public key for encryption or verification and a private key for decryption or signing. Typical applications include key exchange (\eg\/, during a TLS handshake), digital signatures, or the protection of small data items such as cryptographic keys. In contrast to symmetric encryption, asymmetric methods avoid pre-shared secrets and instead rely on computationally hard mathematical problems to ensure that decryption is feasible only with the private key. As it does not benefit to the same degree from parallelism or hardware acceleration, asymmetric encryption is unsuitable for bulk data protection. Instead, modern systems employ hybrid encryption: asymmetric cryptography is used to authenticate users and securely establish or wrap a symmetric session key, which then encrypts the bulk of data. This hybrid design powers TLS, secure file-sharing systems, and cloud key management infrastructures. 

Finally, homomorphic encryption (HE) allows computation on encrypted data without decryption---theoretically attractive for \gl data in use\gr\/ protection in untrusted compute environments. Multiple recent articles examine HE in HPC and cloud contexts. For instance, Meftah et al.\@ (2022) utilise HPC systems to accelerate HE for convolutional neural networks \cite{Meftah2022}. Similarly, Jung et al.\@ (2020) propose algorithms to reduce the computational cost of HE by avoiding expensive multiplications \cite{Jung_2021}. Prantl et al.\@ (2023) demonstrate comparable performance analyses in a practical linear‐regression scenario \cite{Prantl2023}. These results highlight that HE is a promising field, but its current runtime overhead is still orders of magnitude higher than native plaintext processing due to lattice-based arithmetic, ciphertext noise management, and the absence of efficient hardware acceleration. Thus, for large‐scale HPC workloads HE remains impractical and requires significant compute overhead or architectural redesign. That in turn undermines one of the main goals of high‐performance computing: fast throughput and low latency. Hybrid approaches that encrypt part of the data or isolate sensitive segments are presently more feasible. 

In sum, the encryption strategy in a secure computing environment often uses symmetric encryption for bulk data, asymmetric\slash\allowbreak hybrid for key distribution and access, and reserves homomorphic encryption or secure enclaves for highly sensitive or small-volume operations where a performance penalty is acceptable.

\subsection{Hardware Memory Encryption Standards Across CPU and GPU Vendors}
Modern secure computing environments increasingly rely on in-situ memory en\-cryp\-tion to protect against physical attacks (\eg\/, cold-boot, direct memory access, DIMM bus probing) and to harden multi-tenant isolation. Across vendors, this is achieved via an inline AES engine at the memory controller with address-dependent tweaks, ephemeral keys generated in hardware, and per-tenant keys plus integrity metadata. Still, design choices differ in scope (whole-system vs.\@ per-VM), granularity (page-level selection vs.\@ \gl encrypt everything\gr\/), and whether integrity\slash\allowbreak freshness is provided in addition to confidentiality. 

\paragraph{AMD (SME, TSME, SEV, SEV-SNP).} AMD's Secure Memory Encryption (SME) integrates AES-XTS, a tweakable AES- and physical-address-based encryption mode, into the memory controller and encrypts any page marked with the C-bit using a 256-bit key derived by the AMD Secure Processor (SP) at boot. Transparent SME (TSME) extends this to system-wide DRAM encryption where a single key encrypts all memory. Secure Encrypted Virtualization (SEV) builds on SME by assigning each virtual machine its own memory encryption key, preventing the hypervisor or co-tenants from inspecting guest memory. The latest development, SEV-SNP (Secure Nested Paging), adds integrity protection via the Reverse Map Table (RMP) and related mechanisms, thereby blocking remapping and replay attacks (\cite{AMD2020_sev_snp}). This feature adds the attestation mechanisms needed for confidential computing, allowing tenants to verify the platform's hardware and security state before releasing sensitive data---an attestation workflow now standard in major public clouds \cite{AMD2021}. 

\paragraph{Intel.} Intel’s memory encryption technologies span several layers. Total Memory Encryption (TME) provides baseline protection by encrypting all external DRAM with AES-XTS using a single ephemeral key generated inside the CPU during boot; this protects against physical memory attacks and does not require OS changes, but it applies a single key to the entire system. Multi-Key TME (MKTME) generalizes this design by supporting multiple encryption keys and allowing the OS or hypervisor to select keys on a per-page basis, enabling isolation between different workloads or VMs. Trust Domain Extensions (TDX) build on MKTME to create confidential VMs (trust domains) that not only use distinct memory encryption keys, but also add integrity protection mechanisms (such as per-cache-line MACs and anti-replay measures), remote attestation, and a reduced host-side trusted computing base. Together, TME, MKTME, and TDX form a progression from system-wide memory encryption toward a full confidential-computing model comparable in goals to AMD SEV-SNP, albeit through different architectural mechanisms \cite{Intel_TME}. 

\paragraph{Arm (CCA/RME).} Arm’s Confidential Compute Architecture (CCA) introduces Realms: hardware-isolated execution environments that protect code and data even from privileged system software. Realms extend the traditional secure\slash\allowbreak non-secure model with a third, strongly separated domain and rely on the Memory Protection Engine (MPE) to enforce external-memory encryption and, where implemented, integrity or freshness protection. System memory is divided into Physical Address Spaces (PAS), each with their own encryption context, while the MMU\slash\allowbreak SMMU (memory management unit) assigns pages to PASs and ensures that Realm keys are discarded upon teardown. CCA delivers strong architectural isolation with compulsory encrypted-memory boundaries, while moving specific cryptographic mechanisms to dedicated hardware \cite{ARM_RME_2021}. Despite the strong architectural isolation in Arm's CCA, the concept is too new to be applied in current HPC systems which are not yet powered by Armv9 processors (\url{top500.org}). 

\paragraph{NVIDIA.} With the Hopper H100, NVIDIA extended confidential computing to GPU accelerators. The PCIe interconnect between Confidential Virtual Machines (CVMs; \ie\/, VMs running under AMD SEV-SNP or Intel TDX) and the GPU is protected by hardware-enforced authenticated encryption using AES-GCM, ensuring that data traversing the CPU-GPU interface remain confidential and protected against interception or tampering. In addition, the GPU's firmware and boot process are cryptographically measured and remotely attested, allowing a CVM to verify the integrity of the GPU before entrusting it with sensitive workloads. Vendor documentation and early measurements indicate that compute-heavy workloads reach near-native performance, whereas I/O-bound jobs pay a higher overhead due to encrypted paging across the CPU–GPU interconnect \cite{Nvidia2023_H100_confidential_compute}. 

\paragraph{} Functionally, AMD SME\slash\allowbreak SEV-SNP and Intel TME\slash\allowbreak MKTME\slash\allowbreak TDX converge on inline AES with key isolation and CVM-grade attestation, while Arm CCA\slash\allowbreak RME focuses on per-domain memory encryption and a system-level isolation model. NVIDIA extends these principles to accelerators with end-to-end encrypted data transfer between the CPU and the GPU. 
For our use case of secure scientific computing on a multi-tenant HPC system, the AMD platform is the preferred choice because of its native 256-bit hardware encryption engines, the mature ecosystem for running multi-tenant encrypted VMs, its long-established kernel support and simple deployment. 
Although Intel-based CPUs offer more sophisticated attestation mechanisms and will likely catch up in the future, they currently lack operational experience and require a more complex setup. Arm CPUs with Confidential Compute Architecture seem promising, but ARM-based HPC clusters are not yet available, and limitations in compiler ecosystems and application porting present additional barriers for now.

\subsection{Key Management: Secure Key Servers and Life Cycle Management}
\label{key_management}
Encryption alone is not sufficient in secure computing environments. Secure key management is equally important, as cryptographic protection is ruined with compromised keys or improper key handling \cite{Kraehenbuehl2023}. Therefore, modern secure computing environments rely on centralised key manage\-ment services or hardware security modules (HSMs) to organize the generation, storage, rotation, access control, auditing, and retirement of keys. 
Among five evaluated cryptographic key management systems (KMSs), HashiCorp Vault stood out for its suitability to small businesses in terms of features and usability \cite{Kuzminykh2020}. Vault’s enterprise version supports HSM integration, automatic unsealing, and FIPS-compliant seal wrapping (\url{https://developer.hashicorp.com/vault/docs/enterprise/hsm}). On the enterprise side, Thales provides Luna HSMs and cloud HSM services, which can integrate with Vault or other KMSs to ensure keys never leave tamper-resistant hardware and comply with regulatory requirements \cite{Thales2022_vault_luna_hsm_integration}. In multicloud or HPC settings, centralised key life cycle management has been argued to be critical: For example, Thales’s White Paper describes how disparate cloud consoles and BYOK (bring-your-own-key) scenarios complicate governance unless a unified key management layer is used \cite{Thales2022_multicloud_key_management}. 
In the HPC or cloud computing context, deploying such key management services means the compute environment can encrypt data volumes at rest, wrap keys via HSMs, enforce role-based access, log key use, handle key rotation and retirement, and integrate with job schedulers or container orchestration. Without robust key management, even strong encryption is undermined by poor key hygiene.

\section{Secure hardware implementation}

\subsection{File encryption}
\label{file-enc}

File encryption on RAMSES is implemented through the data encryption capabilities of IBM Storage Scale (GPFS) and the Thales CipherTrust Manager (CTM) as a management tool for cryptographic keys. This layout enforces strict access control to data at rest and ensures that encrypted storage areas are accessible exclusively to authorized users. 

\begin{figure}[!hbt]
	\centering
	\includegraphics[trim={0cm 0cm 0cm 0cm},clip,width=0.66\textwidth]{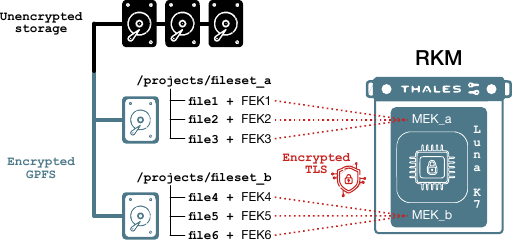} 
	\caption{\textsf{\textbf{GPFS-based file encryption on RAMSES.} File encryption is managed through the GPFS kernel module and the Thales CipherTrust Manager for Remote Key Management (RKM). Two encrypted file sets (fileset\_a, fileset\_b) with three files each are shown. Each file is encrypted via its unique File Encryption Key (FEK1-6), which is stored in the file's extended attributes. FEKs in turn are encrypted by a file set-specific MEK (Master Encryption Key; MEK\_a and MEK\_b) stored safely in the Luna K7 HSM of the Thales CipherTrust Manager. FEK traffic between file sets and the RKM uses an encrypted TLS connection. Access to encrypted file sets is controlled via security policies of the secure partition and corresponding certificates. }}
	\label{file-encryption}
\end{figure}

RAMSES's architecture for the management of file encryption keys is layered and involves three core concepts: the \textit{File Encryption Key}\ (FEK), the \textit{Master Encryption Key}\ (MEK), and \textit{Remote Key Management}\ (RKM). The FEK is a newly generated, unique key; each individual file within an encrypted domain is encrypted by its own FEK. The MEK encrypts the FEKs created during file operations; it is specific to a given file set and securely stored in a hardware security module. Finally, the RKM---implemented via the Thales CipherTrust Manager (CTM)---orchestrates FEK encryption/decryption and secure MEK handling, and protects these critical cryptographic operations with a dedicated, tamper-resistant HSM (Figure~\ref{file-encryption}).

Importantly, file encryption is fully transparent from the user perspective: read, write, copy, and move operations proceed without impacting the user's workflow, as encryption and decryption are performed automatically at the file system level. However, prior to operation, system administrators must define appropriate encryption policies for each encrypted domain by specifying the target file set, the MEK to apply, and the encryption algorithm to be used. Once a file set is designated for encryption, the following sequence of actions is executed transparently in the background when creating or accessing an encrypted file: 

The proprietary GPFS kernel module generates a new random FEK with the CPU's hardware-based random number generator locally on the compute node and encrypts the file with the FEK using the CPU's AES engine. It then establishes a TLS connection to the CTM, authenticates via its X.509 public key certificate, and transmits the plaintext FEK to the CTM over the encrypted TLS channel, requesting that it be wrapped with the MEK specified for the file set. Within the CTM, the request is passed to the Luna K7 HSM which retrieves the MEK from its internal secure storage, uses it to encrypt (wrap) the FEK, and returns the encrypted FEK to the CTM software layer. Thus, all MEK operations occur strictly within the HSM boundary. MEKs never leave the tamper-resistant HSM hardware and do not exist in the memory of the CTM or the host CPU. After the encrypted FEK is sent back to the compute node over TLS, the GPFS kernel module stores the FEK in the file's extended attributes. 

Similar processes take place when accessing an already existing encrypted file. In this case, an encrypted FEK is read from the file's metadata, sent to the CTM over TLS, unwrapped inside the CTM's HSM, and returned in plaintext to the compute node over encrypted TLS, where it is used to decrypt the file. 

Thus, for file encryption to operate correctly, each compute node or virtual machine must be individually configured to authenticate against the CTM via its KMIP (Key Management Interoperability Protocol) interface in order to obtain the required MEK. 
On RAMSES, KMIP certificates for authentication are embedded in the file set-specific VM disk images stored read-only on nodes of the secure partition, \ie\@ the KMIP configuration is already in place upon start-up of the transient VM overlay images. The drawback of this per‑node configuration is substantial administrative overhead, as the numerous virtual machines of a multi-tenant HPC environment all require independent RKM configuration. Moreover, any file set designated for encryption can be accessed only by nodes that have been fully configured in advance; incomplete or missing KMIP configuration prevents decryption and renders the file set inaccessible to that node.

\subsection{Memory encryption}
\label{mem-enc}

\begin{figure}[!hbt]
	\centering
	\includegraphics[trim={0cm 0cm 0cm 0cm},clip,width=0.66\textwidth]{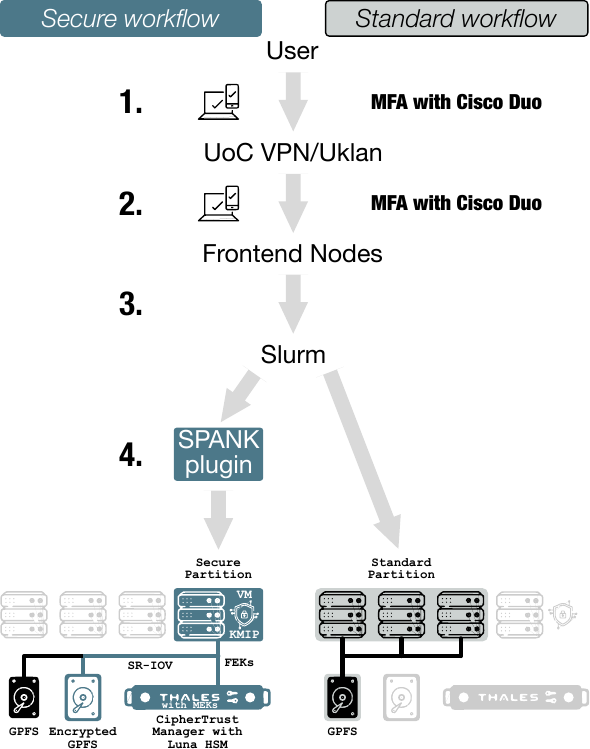} 
	\caption{\textsf{\textbf{Secure vs.\@ traditional HPC workflow on RAMSES.} 
	\textbf{1:} UoC members connect to the UoC network using multi-factor authentication (MFA) based on their IDM credentials and Cisco Duo. \textbf{2:} Registered HPC users login to RAMSES using multi-factor authentication with an SSH key pair and Cisco Duo. \textbf{3:} On the frontend node(s), users submit their jobscript for confidential (left side) or traditional computing (right side) to the Slurm workload manager. \textbf{4:} Left: A dedicated Slurm job parameter activates a custom SPANK plugin essential for confidential computing. After successful verification, a user-specific VM is launched on an SEV-enabled node with active memory encryption. The VM disk image includes a KMIP client certificate, enabling transparent decryption of the associated file set via the Thales CipherTrust Manager and SR-IOV–based GPFS access. FEKs travel over encrypted TLS between the secure partition and the CTM while MEKs stay within the Luna HSM. Upon job completion, the VM is terminated and all runtime state is discarded. Right: Jobs submitted without the confidential-computing parameter follow the standard Slurm-managed execution path. Secure and traditional workloads coexist on the same system and are selected at submission time: running a secure job requires the addition of a single Slurm parameter. Abbreviations: FEK, file encryption key; IDM, identity management system; MEK, master encryption key; SEV, secure encrypted virtualization; SPANK, Slurm Plug-in Architecture for Node and job (K)control; SR-IOV, single root I/O virtualization; Uklan, University of Cologne local area network; UoC, University of Cologne; VM, virtual machine; VPN, virtual private network. }}
	\label{secHPCworkflow}
\end{figure}

The ultimate goal of a secure HPC system is that sensitive data is encrypted at all times, and therefore safe at rest, in transit, and in use. To achieve this, we combined virtualization techniques, sophisticated key management, and the Slurm workload manager \cite{Slurm2023} with dedicated hardware to obtain a coherent workflow for secure data processing (Figure~\ref{secHPCworkflow}). 

To provide a secure HPC environment, we implemented a distinct \gl secure partition\gr\/ on RAMSES, consisting of AMD EPYC Genoa nodes configured with Secure Encrypted Virtualization (SEV) \cite{AMD2020_sev_snp,AMD2021}. These nodes are hardware-identical to standard 192-core SMP nodes but are booted with SEV enabled in the system firmware, allowing the memory controller’s AES engine to transparently encrypt memory pages, \ie\/, encryption is performed entirely in hardware without requiring modifications to applications or the guest operating system. Because SEV activation is controlled at the firmware level, the size of the secure partition can be flexibly adjusted by toggling SEV support and rebooting selected nodes. 

Secure execution is provided through virtual machines managed with a KVM (kernel-based virtual machine; \url{https://linux-kvm.org}) and QEMU (quick emulator; \url{https://www.qemu.org}) software stack. While KVM acts as the in-kernel hypervisor, QEMU supplies the virtual hardware model and device emulation. During VM instantiation, the hypervisor verifies SEV availability using an extended function leaf for hardware-level memory encryption present on AMD processors. If enabled, the hypervisor sets the SEV bit in the VM control structure before issuing VMRUN, causing the guest OS to boot with SEV-protected memory. As a consequence, the hypervisor cannot read encrypted memory contents, protecting VMs booted under SEV from malicious administrators or compromised hypervisors. RAMSES's secure VMs are based on customized Red Hat Enterprise Linux 9.6 images that include KMIP client certificates to retrieve file encryption keys from the Thales CipherTrust Manager. For each user or project, administrators provision dedicated virtual machines with a unique certificate and passphrase for Thales authentication and key retrieval, allowing the user decryption of their corresponding file sets (see section \ref{file-enc}). 

To enable storage access within secure VMs, the InfiniBand interconnect must also be made available in a virtualized form. Since a physical host channel adapter to the InfiniBand network cannot be shared directly across multiple VMs, we employed Single Root I/O Virtualization (SR-IOV) \cite{PCISIG_sriov_2010} on RAMSES's ConnectX\nobreakdash-6 InfiniBand controllers to expose up to \num{127} virtual functions (VFs) from a single InfiniBand card \cite{Nvidia_sriov_doc}. These VFs can be assigned to different VMs running on a single node, providing high-performance access to the parallel file system via IP-over-IB. Custom automation scripts configure and allocate the VFs to VMs on demand. To take into account that one user or group may submit several secure jobs simultaneously, we provide multiple VM disk images for each encrypted file set. 

Finally, confidential workloads are integrated into the cluster workflow through Slurm. A custom SPANK plugin for Slurm (available upon request) verifies that the submitting user is authorized for secure computing. Specifically, our plugin confirms that the corresponding VM exists and that the user is member of a group with an assigned secure VM before placing the job onto a secure node. Slurm then triggers VM instantiation and coordinates the retrieval of the necessary Thales keys before the job is executed. In addition, the Slurm plugin defines custom parameters that allow administrators to access and modify the permanent storage of specific secure VMs within the limits of Slurm jobs. 

Importantly, the secure VMs of RAMSES are transient, \ie\/, they are destroyed upon shutdown. They operate on temporary overlay files derived from immutable backing disk images (read-only). Thus, any modifications made to regular operating system directories during runtime are discarded. Only job output and result files are stored in encrypted form in mounted GPFS directories and will therefore be accessible after job completion. 

In combination, these mechanisms provide end-to-end confidentiality for data at rest (through KMIP and Thales-managed encryption) and data in use (through AMD SEV).

\section{Secure software implementation}

\subsection{User access via multi-factor authentication (MFA)}
RAMSES provides SSH access for unprivileged users via dedicated login nodes. They can be reached directly from within the University of Cologne (UoC) campus network or via the UoC virtual private network (VPN). VPN access requires multi-factor authentication using the password for the UoC identity management system (uniKIM) \cite{unikim} and an additional factor. 

At the University of Cologne, the Cisco Duo application \cite{ciscoduo} is the predominant solution to deliver the second factor. The most common MFA scenario involves authentication through the Duo Mobile App, available for Android, iOS, iPadOS, and watchOS. The second factor can then be a simple touch response to the app's pop up question upon SSH login. Using a mobile device has the advantage of introducing another safety layer created by the device's own access control (PIN, face recognition, or other). Registration of the Duo Mobile App is restricted to the campus network, while external users are subject to an additional identity verification step. As alternatives to a mobile device, we also support hardware tokens and time-based one-time passwords (TOTP) for MFA, the latter only for temporary accounts, such as course accounts. 

\begin{figure}[!hbt]
	\centering
	\includegraphics[trim={0cm 0cm 0cm 0cm},clip,width=0.8\textwidth]{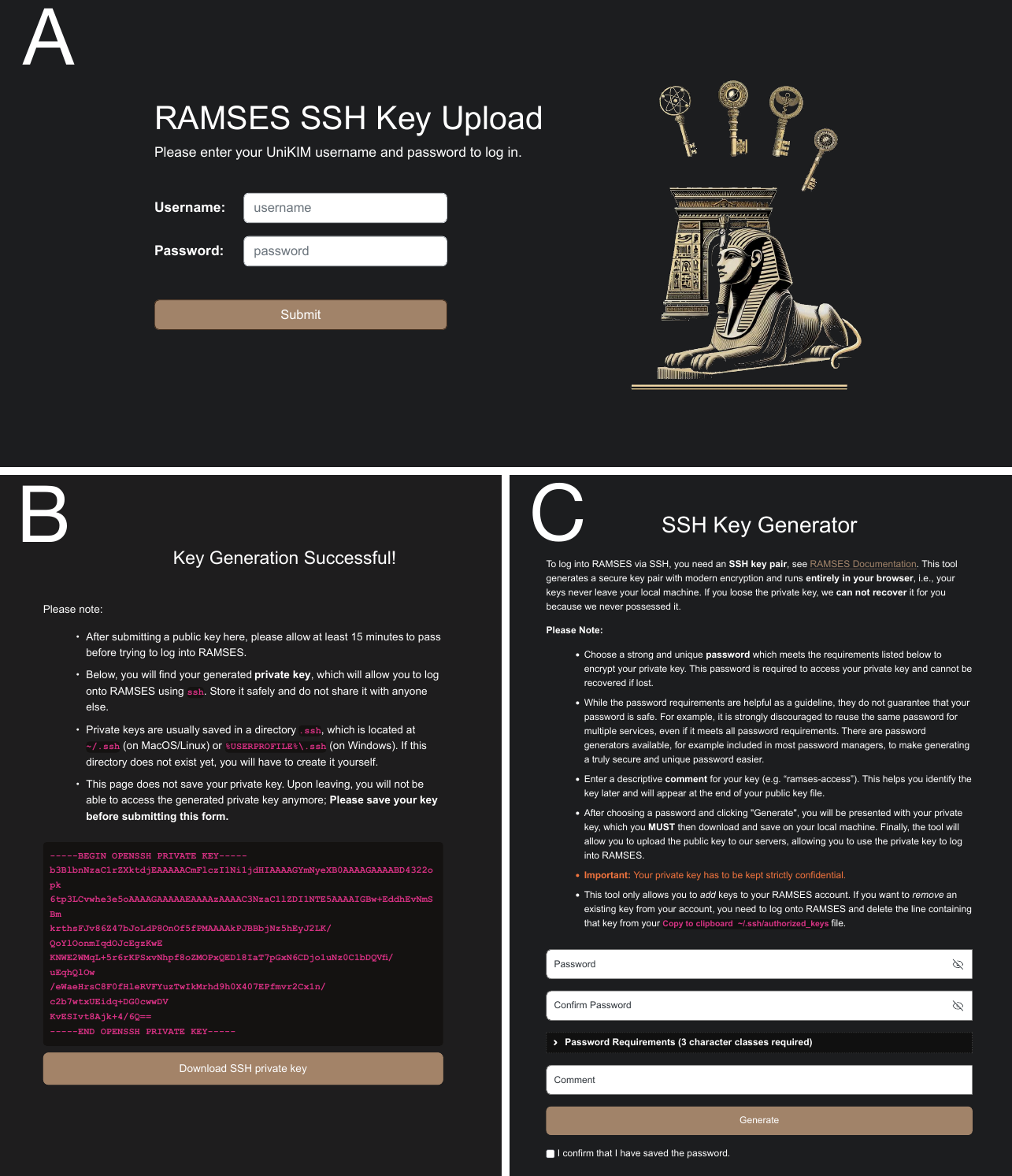} 
	\caption{\textsf{\textbf{The RAMSES SSH key upload website.} 
	\textbf{A:} UoC members with approved RAMSES account can connect to the RAMSES key upload website (\url{https://ramses-umc.itcc.uni-koeln.de/web-sshkey/login}) from within the University network, using their UoC IDM credentials and Cisco Duo. \textbf{B, C:} To improve user experience and security, a JavaScript-Plugin carries out SSH key pair generation and ensures its protection with a strong password. The public key is subsequently copied to the user's \url{~/.ssh/authorized_keys} file on RAMSES while the private key, including its passphrase, must be downloaded to the user's computer. }}
	\label{keyupload}
\end{figure}

Finally, SSH access to the RAMSES login nodes again requires MFA, using SSH key-based authentication \cite{sshkeys} and the specified second factor (Figure~\ref{secHPCworkflow}). To support users in generating SSH keys and to streamline key upload to RAMSES, we deployed a dedicated web service accessible from within the UoC network (Figure~\ref{keyupload}) \cite{upkeys}. The frontend is implemented in Go and incorporates a JavaScript component that performs key generation locally on the client device. The tool supports ED25519 elliptic-curve keys and RSA (Rivest–Shamir–Adleman) keys with a minimum length of 4096 bits and enforces passphrase protection with defined complexity requirements (\eg\/, length, character classes). The backend component provisions the public key via an LDAP (Lightweight Directory Access Protocol) directory service, making it available to RAMSES for authentication. This approach ensures that private keys remain exclusively on the user’s device while simplifying onboarding for users with limited experience in SSH-based access. 

As some users may be unfamiliar with the concept of SSH keys and MFA-related mechanisms, we made comprehensive RAMSES documentation publicly available on GitHub (\url{https://gitlab.git.nrw/uzk-itcc-hpc/itcc-hpc-ramses/-/wikis/home}) and provide beginner-oriented tutorials that describe access and login procedures (link to video).

\subsection{Administrative access and operating system hardening}
Access to the system follows the principles of privileged access management, where physical and logical system components are separated into different security levels, each with different security measures and restrictions. Privileged access is strictly unidirectional: from higher to lower security levels, to prevent escalation of privileges \cite{nist2018_sp1800-18_pam}. 

\begin{figure}[!hbt]
	\centering
	\includegraphics[trim={0cm 0cm 0cm 0cm},clip,width=0.4\textwidth]{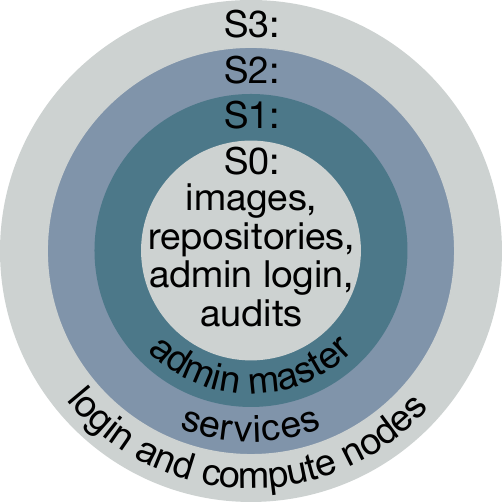} 
	\caption{\textsf{\textbf{RAMSES's security layout.} 
	The physical and logical system components of RAMSES are separated into four security levels, S0 to S3, illustrated as concentric circles. Unprivileged users can only access login and compute nodes of the lowest security level, S3 (outer ring). Access to nodes of higher security levels (S0 to S2) is restricted to privileged accounts and implemented through unidirectionality, hardened entry points, blocking of reverse connections, strict firewall policies, and restrictive SSH configurations. 
    }}
	\label{security_layers}
\end{figure}

As such, nodes and services dedicated to system administration tasks are placed at the higher security level, while nodes directly accessible by unprivileged users, such as login and compute nodes, are at the lowest level (Figure \ref{security_layers}). To enforce strict security measures, administrative nodes are placed behind hardened jump hosts and protected with multi-factor authentication and restrictive firewall policies. 

Internally, administrative services are further differentiated into the following hierarchy: Critical services, including administrative logins, monitoring and logging, image and package repositories occur at the highest security level. The level below contains provisioning and resource management services for login and compute nodes. Administrative nodes running critical services follow unidirectional privileged access, in which management connections must originate from the nodes with higher security levels to the nodes with lower security levels, with all reverse connections blocked. This is enforced by firewall rules and restrictive SSH configurations.

\subsection{Information Security Management System (ISMS)}
To govern the secure handling of sensitive data on RAMSES, the IT~Center Cologne (ITCC) is implementing an Information Security Management System (ISMS) in accordance with the ISO/IEC~27001 standard \cite{ISO27001}. An ISMS is a structured, risk-based framework designed to establish, implement, maintain, and continually improve information security within an organization. 

The ISMS scope covers all organizational and technical aspects of processing and storing sensitive data on RAMSES. It is specifically designed to support research workflows that include special-category data as defined by Article~9 of the General Data Protection Regulation (GDPR), such as human genomic data and data derived from medical imaging procedures \cite{GDPR2016}. Relevant use cases within the scope of the ISMS are rooted primarily in the life sciences and encompass commissioned data processing, research collaborations, diagnostic workflows, and data processing in the context of third-party funded projects. 

The core services provided within the Secure Computing and Storage Environment (SCSE) comprise secure compute resources for the analysis of sensitive data, high-security storage capacity for such data, and operational support for the use of these resources. 

From a technical perspective, data confidentiality within the SCSE is enforced through multiple complementary mechanisms: Data at rest are protected by encrypted file systems (see section \ref{file-enc} and Figure~\ref{secHPCworkflow}); data in use are protected by memory encryption during processing (see section \ref{mem-enc} and Figure~\ref{secHPCworkflow}); and all data transport is secured by encrypted communication channels. These technical measures operate in conjunction with access-control mechanisms and a layered security layout (Figure~\ref{keyupload}, \ref{security_layers}) that are enforced throughout the system. Of note, our organizational implementation of the ISMS together with the comprehensive encryption architecture described in this article go beyond the requirements of the ISO/IEC~27001 standard. 

Information Security concerning the ITCC and its services is handled by the ITCC Security Operations Team. All security incidents and related issues outside the ITCC fall under the responsibility of the UoC Chief Information Security Officer (CISO), who directly answers the UoC rectorate. More information is provided in the UoC guidelines on information security \cite{UzK2025ISL}.

\section{Validation of the approach in the biomedical domain}
To analyse the impact of encryption on job performance, we executed two representative workflows from the life sciences---the primary field with requirements in confidential computing---under various encryption modes. As an I/O-heavy workload, we used the software RepeatMasker \cite{repeatmasker} which identifies repeat elements in a genome sequence. In addition, we used the software BWA-MEM2 \cite{Vasimuddin2019} for aligning short read sequences against a reference genome---a life science workflow with high memory requirements. 

\begin{figure}[!hbt]
	\centering
	\includegraphics[trim={0cm 0cm 0cm 0cm},clip,width=0.7\textwidth]{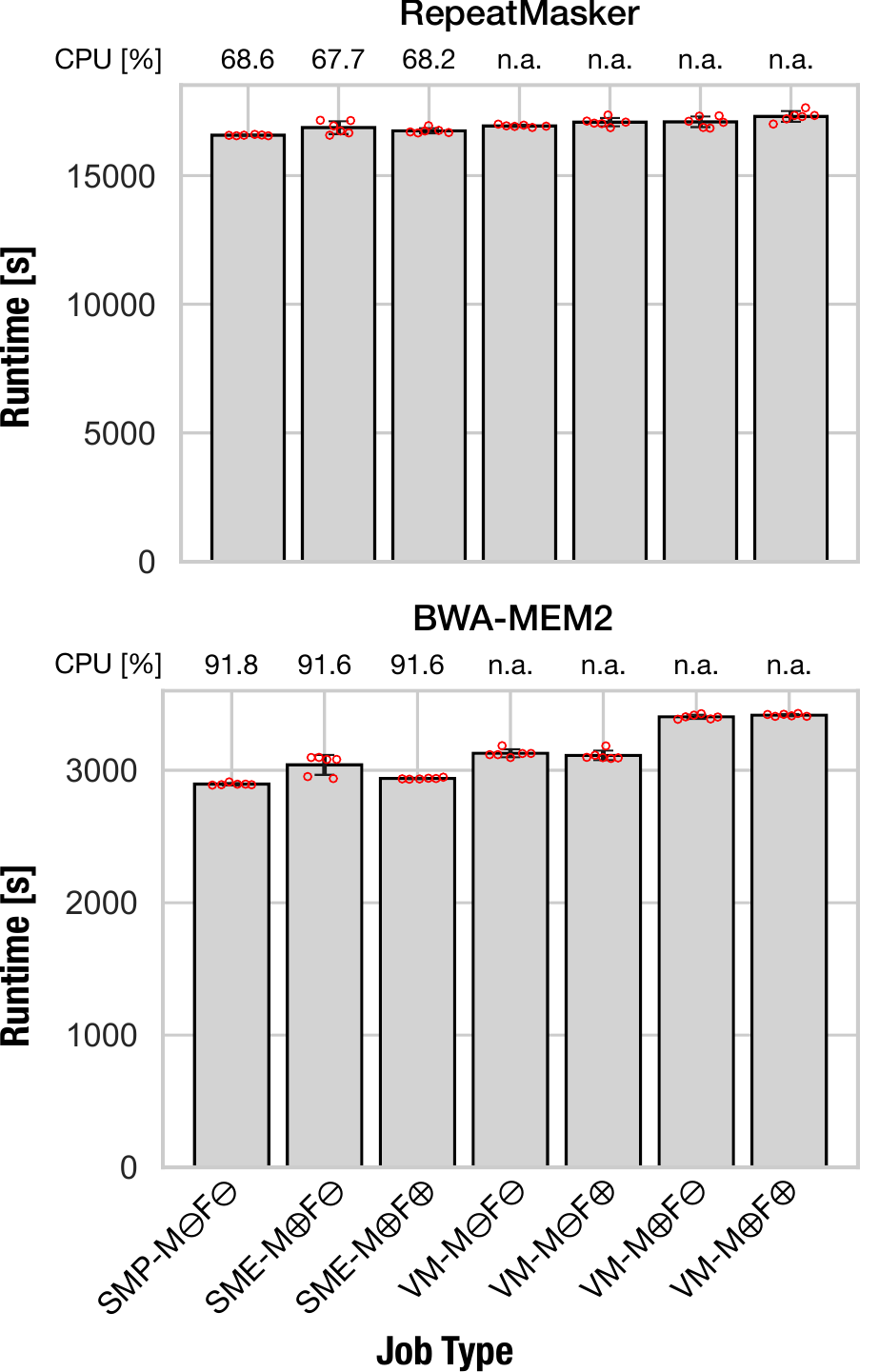} 
	\caption{\textsf{\textbf{Performance of two contrasting workloads under various encryption regimes.} RepeatMasker (top; I/O-heavy) and BWA-MEM2 (bottom; memory-intensive) runs were carried out in seven different configurations with six replicates each: 1) Standard SMP node without memory and without file encryption (SMP-M$\ominus$F$\ominus$); 2) SME-enabled node with memory encryption, without file encryption (SME-M$\oplus$F$\ominus$); 3) SME-enabled node with memory encryption and file encryption (SME-M$\oplus$F$\oplus$); 4) VM on standard SMP node without memory encryption and without file encryption (VM-M$\ominus$F$\ominus$); 5) VM on standard SMP node without memory encryption, but with file encryption (VM-M$\ominus$F$\oplus$); 6) VM on SEV-enabled node with memory encryption, without file encryption (VM-M$\oplus$F$\ominus$); 7) VM on SEV-enabled node with memory and file encryption (VM-M$\oplus$F$\oplus$). Numbers on top denote average CPU efficiency of the respective job type. Red circles represent individual measurements. n.a.: data not available. M: memory encryption; F: file encryption; $\oplus$: enabled; $\ominus$: disabled. }}
	\label{testruns}
\end{figure}

In total, we devised seven different encryption regimes and measured workflow execution times, CPU efficiency, and memory consumption in six replicates per setup and workflow. 
Our analyses reveal that node-exclusive jobs on standard SMP nodes without any form of encryption (native bare-metal) complete fastest in both workflows (Figure~\ref{testruns}, SMP-M$\ominus$F$\ominus$). As expected, running the same jobs in the most elaborate encryption configuration---from within a VM with memory and file encryption (VM-M$\oplus$F$\oplus$)---takes the longest. While the difference in execution times between the least secure (SMP-M$\ominus$F$\ominus$) and the most secure job variant (VM-M$\oplus$F$\oplus$) is \qty{4.4}{\percent} for the I/O-heavy RepeatMasker workflow, this difference increases to \qty{18.0}{\percent} for the memory-intensive BWA-MEM2 workflow (Table~\ref{runtime_summary}). 
Thus, the full encryption setup is associated with a measurable performance decrease that may vary depending on the particular workflow, but retains high-throughput processing capabilities for confidential data. 

Software consisting of alternating parallel and serial stages is common in bio\-informatics pipelines, \eg\ for variant calling \cite{McKenna2010} or genome annotation \cite{Gabriel2024}. We conclude from our RepeatMasker results that workflows with alternating parallel and serial stages experience only minor performance penalties under the full encryption scenario (Figure~\ref{testruns}), possibly because performance drops during serial stages outweigh overheads introduced by encryption and virtualization, as illustrated by the reduction of CPU efficiency from \qtyrange{91}{68}{\percent} between BWA-MEM2 and RepeatMasker runs (Figure~\ref{testruns}). However, to verify if this interpretation is valid in general, it would be necessary to analyse the runtime behaviour of similar workflows in more detail. 

Compared to standard runs, the full encryption scenario introduces overhead from virtualization, memory encryption (SEV), and file system encryption. To isolate the impact of virtualization on job performance, we carried out runs from within a VM on standard SMP nodes lacking memory and file encryption (VM-M$\ominus$F$\ominus$) and compared them to bare-metal performance (SMP-M$\ominus$F$\ominus$). Our results indicate that substantial parts of the observed encryption costs originate from virtualization (Figure~\ref{testruns}, Table~\ref{runtime_summary}). In particular, virtualization accounts for a penalty of \qty{8.1}{\percent}, or \qty{45.0}{\percent} of the overall overhead (\qty{18.0}{\percent}), in the BWA-MEM2 use case and for a penalty of \qty{2.2}{\percent}, or \qty{50.0}{\percent} of the overall overhead (\qty{4.4}{\percent}) in the RepeatMasker test environment, suggesting that the abstraction layer is responsible for roughly half of the observed performance losses (Table~\ref{runtime_summary}). Although an optimized environment with explicit CPU pinning and NUMA-aware memory binding (non-uniform memory access) may reduce overhead, the virtualization costs reported here align well with published data \cite{Giallorenzo2021,Hanussek2021,Kuity2023}. 
Importantly, the results described here represent common workloads from widely used bioinformatics applications that have not been designed for benchmark studies or the operation in an encrypted environment. As such, our setup might involve undesirable algorithm-specific branching patterns, cache behavior, and adaptive processing paths which are less pronounced in synthetic benchmark design. Our workflows might thus emphasize performance differences between job types (\eg\/, with and without encryption) that are less visible in carefully devised benchmark tests, as already put forward in our previous work \cite{Kessler2025}. 

Which type of encryption is more costly, file encryption, or memory encryption? 
In our setting, adding file encryption to a memory-encrypted context has almost no effects on performance compared to otherwise identical runs without file encryption, whether on SME nodes or within a virtual machine. Similarly, enabling memory and file encryption without the use of a dedicated VM has negligible impact on job performance, as demonstrated by runs on memory-encrypted SME nodes (Figure~\ref{testruns}, SME-M$\oplus$F$\oplus$ vs.\@ SME-M$\oplus$F$\ominus$ and SMP-M$\ominus$F$\ominus$). 

In contrast, combining virtualization with file and memory encryption reduces performance by up to \qty{18}{\percent} relative to the unencrypted bare-metal reference. As mentioned above, approximately half of this penalty is attributable to virtualization overhead. The remainder is predominantly caused by memory encryption, as combining file encryption with virtualization in the absence of memory encryption is largely performance-neutral in both workflows (VM-M$\ominus$F$\ominus$ vs.\@ VM-M$\ominus$F$\oplus$). 
This effect is particularly pronounced in the memory-intensive BWA-MEM2 workflow, where enabling memory encryption alone accounts for nearly the full performance reduction observed in the combined encryption scenario (VM-M$\oplus$F$\ominus$ vs.\@ VM-M$\oplus$F$\oplus$), suggesting that memory bandwidth saturation may increase memory encryption overhead in memory-bound workloads. We therefore conclude that memory encryption and virtualization overhead jointly account for the majority of the observed performance penalty, whereas file encryption imposes no significant cost regardless of the execution environment. 

Despite its performance impact, the use of virtual machines for the secure handling of file system keys and certificates is a critical part of our security architecture, and therefore a replacement with less resource-intensive alternatives is currently not meaningful. Nevertheless, the two representative workflows tested here demonstrate that RAMSES is capable of executing production jobs under full encryption with acceptable performance.

\setlength{\tabcolsep}{7pt}
\begin{table}[hbt!]
\centering
\footnotesize
\begin{tabular}{@{}ll S S S@{}}
\toprule
\textbf{Application} & \textbf{Job Type} & \textbf{Mean [s]} & \textbf{Std Dev [s]} & \textbf{Rel.~Runtime [\%]} \\
\midrule
BWA-MEM2 & SMP-M$\ominus$F$\ominus$ & 2895.66 & 8.60 & 100.0 \\\\ \addlinespace[-0.6em]
 & SME-M$\oplus$F$\ominus$ & 3041.57 & 75.05 & 105.0 \\\\ \addlinespace[-0.6em]
 & SME-M$\oplus$F$\oplus$ & 2938.13 & 5.50 & 101.5 \\\\ \addlinespace[-0.6em]
 & VM-M$\ominus$F$\ominus$ & 3128.91 & 29.96 & 108.1 \\\\ \addlinespace[-0.6em]
 & VM-M$\ominus$F$\oplus$ & 3112.62 & 36.10 & 107.5 \\\\ \addlinespace[-0.6em]
 & VM-M$\oplus$F$\ominus$ & 3404.14 & 16.24 & 117.6 \\\\ \addlinespace[-0.6em]
 & VM-M$\oplus$F$\oplus$ & 3416.86 & 9.18 & 118.0 \\\\ \addlinespace[-0.2em]
RepeatMasker & SMP-M$\ominus$F$\ominus$ & 16563.32 & 19.45 & 100.0 \\\\ \addlinespace[-0.6em]
 & SME-M$\oplus$F$\ominus$ & 16859.92 & 247.82 & 101.8 \\\\ \addlinespace[-0.6em]
 & SME-M$\oplus$F$\oplus$ & 16737.50 & 103.17 & 101.1 \\\\ \addlinespace[-0.6em]
 & VM-M$\ominus$F$\ominus$ & 16927.98 & 44.08 & 102.2 \\\\ \addlinespace[-0.6em]
 & VM-M$\ominus$F$\oplus$ & 17073.85 & 159.94 & 103.1 \\\\ \addlinespace[-0.6em]
 & VM-M$\oplus$F$\ominus$ & 17083.98 & 205.83 & 103.1 \\\\ \addlinespace[-0.6em]
 & VM-M$\oplus$F$\oplus$ & 17296.67 & 206.23 & 104.4 \\\\ \addlinespace[-0.6em]
\bottomrule
\end{tabular}
\caption{\textsf{\textbf{Runtime summary of two workloads under various encryption regimes (job types).} Relative runtime of the bare-metal SMP node (SMP-M$\ominus$F$\ominus$) is set to \qty{100}{\percent}. M: memory encryption; F: file encryption; $\oplus$: enabled; $\ominus$: disabled. }}
\label{runtime_summary}
\end{table}

\section{Conclusions}
The RAMSES platform demonstrates that it is possible to design HPC environments which meet the most stringent data protection requirements without sacrificing the scientific productivity that HPC users expect. By integrating three complementary layers---hardware‑based memory encryption (AMD SEV‑SNP) \cite{AMD2020_sev_snp}, end‑to‑end file system encryption organized through the Thales CipherTrust Manager (FIPS‑validated HSM) \cite{Thales2022_multicloud_key_management,Thales2022_vault_luna_hsm_integration}, and a secure identity‑and‑access framework built on multi‑factor authentication and hardened zones for privileged acces---RAM\-SES enforces end-to-end encryption across the full data life cycle: at rest, in transit, and in use. 

\paragraph{Comparison with cloud‑based confidential computing.}
Confidential computing services in the public cloud (\eg\/, on AMD SEV‑SNP or Intel TDX instances) provide strong isolation but typically require users to migrate data to external environments, incurring additional legal and logistical overheads for regulated datasets. In addition, cloud users are dependent on what the cloud provider supports in terms of audit logs and monitoring interfaces, which may be limited. In contrast, RAMSES retains data within the university’s trusted boundary, complying with GDPR, ISO/IEC~27001, and NIST SP 800‑53 requirements while offering the same hardware‑level guarantees as cloud providers \cite{CCC2022Outreach}. Moreover, the on‑premises deployment permits detailed auditing, custom policy enforcement, and physical access control that are often unavailable or cost‑prohibitive in public clouds, especially as our services are provided to users free of charge. Sensitive data never leave the university's trusted infrastructure, ensuring full institutional control and preventing risks associated with third-party data custody. 

\paragraph{Advances over existing secure HPC architectures.}
Compared with certified HPC centres that rely solely on organisational measures such as ISO~27001 certification, or with Trusted Research Environments that restrict protection to data at rest and in transit \cite{tre_green_paper2020,Kavianpour2022,EGI2024_tre_landscape}, RAMSES extends the security envelope to encompass data in use through hardware-enforced memory encryption---a capability absent from previously reported secure HPC prototypes, including the GWDG workflow \cite{Nolte2022}. 
By implementing memory encryption at the hardware level and automating key management through the Thales CipherTrust Manager, RAMSES addresses this gap without compromising usability: a single Slurm parameter activates a secure execution environment. The necessary virtual machine and cryptographic keys are provisioned automatically, and the user's workflow requires no modifications. 

\paragraph{Performance impact and scalability.}
Our benchmark suite---representing both I/O‑bound (RepeatMasker) and memory‑bound (BWA‑MEM2) bioinformatics pipelines---shows that the full security stack incurs a modest overall runtime penalty (\qtyrange{4}{18}{\percent}). Detailed analysis reveals that roughly half of this overhead originates from virtualization, while memory encryption contributes the remainder in memory‑intensive workloads; file‑system encryption adds negligible cost. These figures align with recent studies on confidential‑computing overheads \cite{Giallorenzo2021,Hanussek2021,Kuity2023} and confirm that RAMSES can support production‑scale scientific workloads with acceptable throughput. 

\paragraph{Future work.}
To further narrow the performance gap and broaden the security envelope, we plan to:

\begin{itemize}
 \item \textbf{Automate VM handling:} automate VM provisioning to reduce administrative overhead and implement secure and reliable VM logging mechanisms. 
 \item \textbf{Reduce virtualization overhead:} investigate lightweight container‑based isolation (\eg\/, Kata Containers \cite{KataContainers2024}) and NUMA‑aware VM tuning to minimise the VM‑related penalty while preserving the secure key‑management model.
 \item \textbf{Extend confidential computing to accelerators:} integrate confidential compute capabilities of both NVIDIA and AMD GPUs, thereby protecting sensitive GPU-resident data, while taking advantage of multi-tenant GPU sharing.  
\end{itemize}

\paragraph{Conclusion.}
RAMSES establishes a practical blueprint for next‑generation secure HPC: a tightly integrated stack that satisfies regulatory mandates, matches the security guarantees of cloud confidential computing, and delivers performance competitive with traditional HPC systems. By making our source code available upon request, we enable interested sites to evaluate, adapt, and build upon this architecture, supporting the emergence of a future class of HPC platforms that can handle sensitive scientific data without compromise.

\section{Materials and Methods}
\subsection{I/O-intensive workloads with encryption}
To analyse the influence of encryption on job performance, we executed typical workflows from the life sciences under various encryption schemes and compared the outcome to standard runs without encryption. All measurements were conducted in six replicates on exclusively reserved compute nodes to avoid interference from unrelated workloads. 
As an example of I/O-heavy workloads, we used RepeatMasker version 4.1.5 \cite{repeatmasker} to identify repeat elements in the genome sequence of the great tinamou, \textit{Tinamus major}\/, with the help of a repeat library that we previously created \cite{Palitzsch2025}. For this purpose, we used the tinamou reference genome assembly ASM3246655v1 (\url{https://www.ncbi.nlm.nih.gov/datasets/genome/GCA_032466555.1/}) and a \textit{de novo}\ library containing \num{13712} distinct repeat elements, collected from \num{25} bird species using RepeatModeler \cite{Flynn2020} and Mitetracker \cite{Crescente2018}. RepeatMasker runs were carried out in SMP (symmetric multi-processor) mode using \num{12} cores and \qty{20}{GB} main memory, resulting in runtimes of approximately \qty{280}{\min}. 

We performed seven identical runs for each of the following encryption configurations: 1) Standard SMP node without memory and without file encryption; 2) SME-enabled node with memory encryption, without file encryption; 3) SME-enabled node with memory encryption and file encryption; 4) Standard SMP node plus VM, without memory and without file encryption; 5) Standard SMP node plus VM, without memory, but with file encryption; 6) SEV-enabled node plus VM with memory encryption, without file encryption. 7) SEV-enabled node plus VM with memory and file encryption. We recorded the runtimes of the jobs (results of the \texttt{time} and \texttt{seff <jobid>} command) and plotted their means and standard deviations using jupyter lab 4.5.2 (\url{https://jupyter.org/}) and a standard python stack for scientific computing (matplotlib 3.10.8; numpy 2.4.1; pandas 3.0.0; plotly 6.5.2; seaborn 0.13.2; scipy 1.17.0). 

\subsection{Memory-bound workloads with encryption}
We used BWA-MEM2 version 2.2.1 \cite{Vasimuddin2019} to align Illumina short read sequences against a reference genome assembly---a typical workflow from the life sciences with high memory but low I/O requirements. BWA-MEM2 runs were carried out in SMP (symmetric multi-processor) mode using \num{48} cores and \qty{200}{GB} main memory, resulting in runtimes of approximately \qty{50}{\min}. In each run, we mapped Illumina HiSeq X Ten paired end sequencing reads (downloaded from the European Nucleotide Archive; accession: SRR7733443 with \num{378590104} reads) against the \textit{Homo sapiens}\ GRCh38 primary genome assembly (\url{https://ftp.ensembl.org/pub/release-115/fasta/homo_sapiens/dna/Homo_sapiens.GRCh38.dna.primary_assembly.fa.gz}) using standard parameters (\url{https://github.com/bwa-mem2/bwa-mem2?tab=readme-ov-file}). Like above, analyses were carried out under seven different encryption layouts.

\section{Code availability}
\begin{itemize}
    \item The source code for the custom Slurm SPANK plugin and VM provisioning is available to academic institutions upon request. 
    \item The source code for the RAMSES key upload website is available to academic institutions upon request. 
\end{itemize}

\section{Acknowledgements}
We thank the Deutsche Forschungsgemeinschaft (DFG, German Research Foun\-dation)---project number INST 216/512-1 FUGG, the MKW (Ministry of Culture and Science of the German state of North Rhine-Westphalia---project number 124-4-01.03.02), and the BMFTR (Federal Ministry of Research, Technology and Space) for funding; Oliver Kaiser and Lion Rexhepi for programming the JavaScript component of the RAMSES SSH key upload web service; Dr.~Anja Kootz (Institute for African Studies and Egyptology, UoC) for help with RAMSES's hieroglyphic artwork; Nikolai Wansart, Aileen Diefenthal, and Mauritius Hoevels for help with the RAMSES video tutorials.

\section{Author's contributions}
\textbf{Viktor Achter:} conceptualization, funding acquisition, writing---review and editing. 
\textbf{Stefan Borowski:} software and hardware implementation, writing---review and editing. 
\textbf{Michael Commer:} writing---review and editing. 
\textbf{Peter Heger:} formal analysis, investigation, methodology, validation, visualization, writing---original draft, writing---review and editing. 
\textbf{Lech Nieroda:} conceptualization, investigation, methodology, software and hardware implementation, writing---review and editing. 
\textbf{Roland Pabel:} software and hardware implementation. 
\textbf{Martin Peifer:} Resources. 
\textbf{Christoph Stollwerk:} software and hardware implementation, writing---review and editing. 
\textbf{Kamil Tokmakov:} writing---review and editing. 
\textbf{Stefan Wesner:} funding acquisition, writing---review and editing.


\bibliographystyle{splncs03}
\bibliography{next-gen-hpc-sec}


%
%


\end{document}